\documentclass[aps,pra,12pt,onecolumn,notitlepage,groupedaddress,nofootinbib]{revtex4-2}

\usepackage[utf8]{inputenc}
\usepackage[T1]{fontenc}
\usepackage{lmodern}
\usepackage{amsmath,amssymb,amsthm,mathtools,bm}
\usepackage{hyperref}
\usepackage{booktabs}
\usepackage{graphicx}
\usepackage[a4paper,top=22mm,bottom=22mm,left=27mm,right=27mm]{geometry}
\usepackage{placeins}
\usepackage{needspace}

\hypersetup{colorlinks=true,linkcolor=blue,citecolor=blue,urlcolor=blue}

\newtheorem{theorem}{Theorem}
\newtheorem{proposition}[theorem]{Proposition}
\newtheorem{lemma}[theorem]{Lemma}
\newtheorem{corollary}[theorem]{Corollary}
\theoremstyle{remark}

\newcommand{\CP}{\mathrm{CP}}
\newcommand{\coCP}{\mathrm{coCP}}
\newcommand{\DEC}{\mathrm{DEC}}

\newcommand{\M}{\mathrm M}
\newcommand{\D}{\mathbb D}

\begin{document}
\title{Phase-dependent atomicity and PPT edge states for cyclic-sign covariant qutrit maps}
\author{Adam Rutkowski}
\email{adam.rutkowski@ug.edu.pl}
\affiliation{Institute of Theoretical Physics and Astrophysics,
Faculty of Mathematics, Physics and Informatics, University of Gdańsk,
80-952 Gdańsk, Poland}
\date{November 2026}

\begin{abstract}
Decomposable positive maps cannot detect entangled states with
positive partial transpose (PPT). We obtain a necessary and sufficient
criterion for decomposability in the entire seven-real-parameter class
of Hermiticity-preserving cyclic-sign covariant maps on $M_3$.
The criterion reduces to a one-variable inequality and, for positive
maps in this class, characterizes indecomposability exactly.
We apply it to a phase-dependent family obtained by adding a completely
positive map to the Choi map. Every map in this family is positive,
and we find the exact phase-dependent threshold for indecomposability
and atomicity. At three phase values, the maps remain indecomposable
for every finite deformation strength. Each indecomposable map
gives an indecomposable entanglement witness through its Choi matrix.
We construct PPT edge states of birank $(5,5)$ detected by these
witnesses throughout the indecomposable region.
\end{abstract}
\maketitle

\section{Introduction}
\label{sec:intro}

To detect entangled states with positive partial transpose (PPT),
one needs positive maps that are not decomposable
\cite{Peres1996,Horodecki1997PPT,Choi1975}.
A decomposable map is a sum of a completely positive and a completely
copositive map. Such maps cannot detect PPT entanglement
\cite{Stormer1963,Woronowicz1976}. Finding exact conditions for
indecomposability is therefore an important problem.

Exact decomposability criteria are known for generalized Choi maps
and several extensions
\cite{ChoKyeLee1992,ChruscinskiMarciniakRutkowski2018,
BeraScalaSarbickiChruscinski2025}. Recent work has also considered
sparse positive maps
\cite{PoderiniEtAl2026Sparse,Sacchi2026Sparse,Sacchi2026Competing}.
Symmetry offers another approach through covariant maps and invariant
PPT operators
\cite{SinghNechita2021,KopszakMozrzymasStudzinski2020,
GulatiNechitaSingh2025,ParkEtAl2024,Szczygielski2026}.

We study cyclic-sign covariant maps on $M_3=M_3(\mathbb C)$.
Their symmetry is associated with the tetrahedral group $A_4$,
and the Hermiticity-preserving maps form a known
seven-real-parameter class within the diagonal-orthogonal-covariant
(DOC) and local cyclic-sign invariant (LCSI) settings.
Our main result is a necessary and sufficient condition for
decomposability in this entire class. It reduces the problem to an
inequality in one real variable. Positivity is not assumed in this
criterion. For maps that are positive, failure of the criterion
is equivalent to indecomposability.

We then apply the criterion to a phase-dependent deformation of the
Choi map. All maps in this family are positive, so we obtain an exact
threshold for indecomposability. For three phase values, the maps
remain indecomposable for every finite deformation strength.
By the Choi--Jamio{\l}kowski correspondence, the indecomposable maps
give entanglement witnesses. We construct explicit PPT edge states
of birank $(5,5)$ detected by these witnesses throughout the
indecomposable region. In dimension three, indecomposability of a
positive map also implies atomicity \cite{YangLeungTang2016},
so the threshold determines atomicity as well.

\section{Cyclic-sign covariant maps}
\label{sec:family}

The cyclic permutation and a diagonal sign change are represented by
\begin{equation}
\label{eq:generators}
R=
\begin{pmatrix}
0&0&1\\
1&0&0\\
0&1&0
\end{pmatrix},
\qquad
S=
\begin{pmatrix}
1&0&0\\
0&-1&0\\
0&0&-1
\end{pmatrix}.
\end{equation}

The matrices $R$ and $S$ generate a group isomorphic to
$V_4\rtimes C_3\simeq A_4$, where $V_4$ is the Klein four group
and $C_3$ is the cyclic group of order three.  This is also the $d=3$ cyclic-sign symmetry studied in
\cite{GulatiNechitaSingh2025}.  A map $\Phi$ is cyclic-sign covariant if

\begin{equation}
\label{eq:covariance}
 \Phi(UXU^\dagger)=U\Phi(X)U^\dagger
\end{equation}
for every $U$ in the group generated by $R$ and $S$.

It is enough to check \eqref{eq:covariance} for $R$ and $S$, since
they generate the group.

We use the Choi convention
\begin{equation}
\label{eq:Choi-def}
 C_\Phi=\sum_{i,j=1}^3 E_{ij}\otimes\Phi(E_{ij}),
 \qquad E_{ij}=|i\rangle\langle j|.
\end{equation}
Here $E_{ij}$ denotes the standard matrix unit.

\begin{proposition}
\label{prop:family}
Every Hermiticity-preserving cyclic-sign covariant map
$\Phi:\M_3\to\M_3$, i.e.
$\Phi(X^\dagger)=\Phi(X)^\dagger$, has the form
\begin{equation}
\label{eq:family}
\Phi_{a,b,c;u,v}(X)=
\begin{pmatrix}
a x_{11}+b x_{22}+c x_{33}
&u x_{12}+\bar v x_{21}
&\bar u x_{13}+v x_{31}
\\
v x_{12}+\bar u x_{21}
&c x_{11}+a x_{22}+b x_{33}
&u x_{23}+\bar v x_{32}
\\
\bar v x_{13}+u x_{31}
&v x_{23}+\bar u x_{32}
&b x_{11}+c x_{22}+a x_{33}
\end{pmatrix},
\end{equation}
where $a,b,c\in\mathbb R$ and $u,v\in\mathbb C$.
Conversely, every map \eqref{eq:family} is cyclic-sign covariant.
\end{proposition}

\begin{proof}

Covariance under the diagonal sign matrices separates diagonal and
off-diagonal entries.  In each $(i,j)$ off-diagonal entry, only
$x_{ij}$ and $x_{ji}$ are allowed.  Covariance under $R$ relates the
coefficients on cyclic orbits.  Hermiticity preservation fixes the
complex conjugations.  These conditions give \eqref{eq:family}.
Conversely, direct substitution shows covariance under $R$ and $S$,
and therefore under the group they generate.

\end{proof}

\Needspace{18\baselineskip}
We use the product basis in lexicographic order, with
$|ij\rangle=|i\rangle\otimes|j\rangle$.
The Choi matrix of \eqref{eq:family}, defined by \eqref{eq:Choi-def}, is
\begingroup
\setlength{\arraycolsep}{3.4pt}
\renewcommand{\arraystretch}{1.08}
\begin{equation}
\label{eq:choi-matrix-full}
C_{\Phi_{a,b,c;u,v}}=
\left(\begin{array}{ccc|ccc|ccc}
a&0&0&0&u&0&0&0&\bar u\\
0&c&0&v&0&0&0&0&0\\
0&0&b&0&0&0&\bar v&0&0\\ \hline
0&\bar v&0&b&0&0&0&0&0\\
\bar u&0&0&0&a&0&0&0&u\\
0&0&0&0&0&c&0&v&0\\ \hline
0&0&v&0&0&0&c&0&0\\
0&0&0&0&0&\bar v&0&b&0\\
u&0&0&0&\bar u&0&0&0&a
\end{array}\right).
\end{equation}
\endgroup

The lines in \eqref{eq:choi-matrix-full} separate the three
$3\times3$ blocks associated with the first tensor factor.
The matrix preserves the subspaces spanned by
$|11\rangle,|22\rangle,|33\rangle$ and by the three pairs
$(|12\rangle,|21\rangle)$, $(|23\rangle,|32\rangle)$ and
$(|31\rangle,|13\rangle)$. In this order, it has the block form

\begin{equation}
\label{eq:choi-blocks}
C_\Phi\simeq A(a,u)\oplus B(b,c,v)^{\oplus3},
\end{equation}
with
\begin{equation}
\label{eq:block-defs}
A(a,u)=
\begin{pmatrix}
a&u&\bar u\\
\bar u&a&u\\
u&\bar u&a
\end{pmatrix},\qquad
B(b,c,v)=
\begin{pmatrix}
c&v\\
\bar v&b
\end{pmatrix}.
\end{equation}
By Choi's criterion \cite{Choi1975CP}, complete positivity is equivalent
to $C_\Phi\ge0$, i.e. to positivity of both blocks in
\eqref{eq:block-defs}.  Positivity of these blocks gives

\begin{equation}
\label{eq:CPconditions}
\begin{split}
&a+2\operatorname{Re} u\ge0,\\
&a-\operatorname{Re} u-\sqrt3\,\operatorname{Im} u\ge0,\qquad
a-\operatorname{Re} u+\sqrt3\,\operatorname{Im} u\ge0,\\
&b\ge0,\qquad c\ge0,\qquad |v|^2\le bc.
\end{split}
\end{equation}

For computational convenience, throughout this work we take the
partial transpose with respect to the first subsystem, namely
\begin{equation}
\label{eq:PT}
 X^\Gamma=(T\otimes\operatorname{id})(X),
\end{equation}
where $T$ denotes matrix transposition in the fixed computational basis.
For the parametrization \eqref{eq:choi-matrix-full}, this gives
\begin{equation}
 (a,b,c;u,v)\longmapsto(a,b,c;\bar v,\bar u).
\end{equation}

\section{Decomposability criterion}
\label{sec:dec}

For convenience, we represent a completely copositive (co-CP) map
as $\Psi\circ T$, where $\Psi$ is completely positive. Thus a map is decomposable  when
\begin{equation}
\label{eq:decomp-Choi}
 \Phi=\Phi_1+\Phi_2\circ T
 \quad\Longleftrightarrow\quad
 C_\Phi=P+Q^\Gamma,\qquad P,Q\ge0,
\end{equation}
where $\Phi_1,\Phi_2$ are completely positive,
$P=C_{\Phi_1}$, and $Q=C_{\Phi_2}$. 

Let
\begin{equation}
\label{eq:triangle}
 \Delta=\operatorname{conv}\{1,\omega,\omega^2\},\qquad
 \omega=e^{2\pi i/3},\qquad
 \mathbb D=\{z\in\mathbb C:|z|\le1\}.
\end{equation}
For $A,B,C\in\mathbb R$ and $U,V\in\mathbb C$, denote by
$\rho(A,B,C;U,V)$ the Hermitian matrix obtained from
\eqref{eq:choi-matrix-full} by replacing $(a,b,c;u,v)$ with
$(A,B,C;U,V)$. It has the same block structure as the Choi matrix.

\begin{lemma}
\label{lem:PPT}
The matrix $\rho(A,B,C;U,V)$ is PPT if and only if
\begin{equation}
\label{eq:PPT}
 A,B,C\ge0,\qquad U,V\in A\Delta,\qquad
 |U|,|V|\le\sqrt{BC}.
\end{equation}
\end{lemma}

\begin{proof}
The $3\times3$ block has eigenvalues
$A+2\operatorname{Re}(U\omega^j)$, $j=0,1,2$. Hence it is positive
if and only if $A\ge0$ and $U\in A\Delta$.
Together with the $2\times2$ blocks of \eqref{eq:choi-blocks},
positivity of $\rho$ is equivalent to
$A,B,C\ge0$, $U\in A\Delta$ and $|V|\le\sqrt{BC}$.
Partial transposition sends $(U,V)$ to $(\bar V,\bar U)$;
see \eqref{eq:PT}. Thus $\rho^\Gamma\ge0$ adds
$\bar V\in A\Delta$ and $|U|\le\sqrt{BC}$.
Since $\bar\Delta=\Delta$, these conditions are equivalent to
\eqref{eq:PPT}.
\end{proof}

By PPT--decomposable duality, a Hermiticity-preserving map is
decomposable if and only if its Choi matrix has non-negative trace
against every PPT operator \cite{SinghNechita2021}.
For a cyclic-sign covariant map it is enough to consider the PPT
matrices above. Indeed, let $G=\langle R,S\rangle$.
Averaging any PPT matrix $\rho$ over the group gives
\[
 \frac1{|G|}\sum_{g\in G}(\bar g\otimes g)\rho
 (\bar g\otimes g)^\dagger.
\]
This average is still PPT and has the same trace against $C_\Phi$,
since $C_\Phi$ is invariant under $\bar g\otimes g$.
The Choi correspondence and Proposition~\ref{prop:family} show that
every invariant Hermitian matrix has the form
$\rho(A,B,C;U,V)$. We therefore need only the pairing
\begin{equation}
\label{eq:pairing}
 \frac13\operatorname{Tr}\!\left[
 C_{\Phi_{a,b,c;u,v}}\rho(A,B,C;U,V)\right]
 =aA+bB+cC+2\operatorname{Re}(\bar uU)
 +2\operatorname{Re}(\bar vV).
\end{equation}

For $r\ge0$ and $z\in\mathbb C$, define
\begin{equation}
\label{eq:m}
 m_r(z)=\min_{\xi\in\Delta,\;|\xi|\le r}
 \operatorname{Re}(\bar z\xi).
\end{equation}
The next theorem gives an exact condition in terms of this function.

\begin{theorem}
\label{thm:DEC}

For a Hermiticity-preserving map $\Phi_{a,b,c;u,v}$, we have

\begin{equation}
\label{eq:DEC}
 \Phi_{a,b,c;u,v}\in\DEC
 \quad\Longleftrightarrow\quad
 \begin{cases}
 a,b,c\ge0,\\
 a+2r\sqrt{bc}+2m_r(u)+2m_r(v)\ge0\quad\text{for all }r\ge0.
 \end{cases}
\end{equation}
\end{theorem}

\begin{proof}
Every decomposable map is positive.  Evaluating
$\Phi_{a,b,c;u,v}$ on $E_{11},E_{22},E_{33}$ gives $a,b,c\ge0$.

By the duality and averaging argument above, it is enough to check
that \eqref{eq:pairing} is non-negative for all matrices satisfying
\eqref{eq:PPT}.  If $A=0$, \eqref{eq:PPT} gives $U=V=0$, and the trace is
$bB+cC\ge0$.  For $A>0$, we may divide by $A$ and set $A=1$.
Write $r=\sqrt{BC}$.  For fixed $r\ge0$,

\begin{equation}
 \inf_{B,C\ge0,\ BC=r^2}(bB+cC)=2r\sqrt{bc};
\end{equation}
this includes $bc=0$, where the infimum need not be attained.

Since $U$ and $V$ vary independently in $\Delta\cap r\D$, their
minimum contributions are $m_r(u)$ and $m_r(v)$.
This gives exactly the condition in \eqref{eq:DEC}.

\end{proof}

Theorem~\ref{thm:DEC} applies to every Hermiticity-preserving map
in the seven-parameter class; no positivity assumption is needed.
For a positive map in this class, failure of \eqref{eq:DEC}
is equivalent to indecomposability. Without positivity,
failure of the condition only rules out decomposability.

We now calculate $m_r$.  For $z\neq0$, let
$\delta(z)\in[0,\pi/3]$ denote the angular distance of
$-z/|z|$ from the nearest vertex of $\Delta$, and define
\begin{equation}
 r_0(\delta)=\frac{1}{2\cos(\pi/3-\delta)}.
\end{equation}

\begin{proposition}
\label{prop:m}
For $z\neq0$,
\begin{equation}
 m_r(z)=-|z|\,h_r(\delta(z)),
\end{equation}
where
\begin{equation}
\label{eq:h}
h_r(\delta)=
\begin{cases}
r,
&0\le r\le r_0(\delta),\\[2mm]
\dfrac12\cos\!\left(\dfrac\pi3-\delta\right)
+\sqrt{r^2-\dfrac14}\,
\cos\!\left(\delta+\dfrac\pi6\right),
&r_0(\delta)\le r\le1,\\[3mm]
\cos\delta,
&r\ge1.
\end{cases}
\end{equation}
For $z=0$, $m_r(0)=0$.
\end{proposition}

The proof is given in Appendix~\ref{app:geometry}.
The result also agrees with the known criterion for generalized Choi
maps \cite{ChruscinskiMarciniakRutkowski2018}.  Indeed, for
$u=-1$ and $v=0$,

\begin{equation}
 m_r(-1)=-\min\{r,1\},
\end{equation}
and Theorem~\ref{thm:DEC} reduces to
\begin{equation}
 a+2\sqrt{bc}\ge2.
\end{equation}
This is the standard condition for this family.

\section{Phase-dependent indecomposability and atomicity}
\label{sec:phase}

The criterion above does not by itself guarantee positivity.
To obtain explicit indecomposable positive maps, consider
\begin{equation}
\label{eq:phasefamily}
 \Phi_{\lambda,\theta}
 =\Phi_{1,0,1;-1,0}
 +\lambda\Phi_{0,1,1;0,e^{i\theta}},
 \qquad \lambda\ge0.
\end{equation}
The first term is the positive Choi map, and the second is
completely positive by \eqref{eq:CPconditions}. Thus
$\Phi_{\lambda,\theta}$ is positive for every $\lambda\ge0$ and
every $\theta$. In this family, failure of the criterion
\eqref{eq:DEC} therefore means indecomposability.

In \eqref{eq:family}, this choice corresponds to
$(a,b,c;u,v)=(1,\lambda,1+\lambda;-1,\lambda e^{i\theta})$.
For brevity, write $\delta(\theta)=\delta(e^{i\theta})$.  Then

\begin{equation}
\label{eq:q}
 q(\theta)
 :=
 -\min_{k=0,1,2}\operatorname{Re}(e^{-i\theta}\omega^k)
 =
 \cos\delta(\theta)
 \in[1/2,1].
\end{equation}

\begin{theorem}
\label{thm:threshold}
Define the phase-dependent threshold
\begin{equation}
\label{eq:lambda}
\lambda_{\rm dec}(\theta)=
\begin{cases}
\displaystyle
\frac{\sqrt{\dfrac{2}{1-q(\theta)}}-1}{2(1+q(\theta))},
& q(\theta)<1,\\[3mm]
+\infty, & q(\theta)=1.
\end{cases}
\end{equation}
For every finite $\lambda\ge0$, the positive map
$\Phi_{\lambda,\theta}$ is indecomposable if and only if
\begin{equation}
\label{eq:indec-interval}
0\le\lambda<\lambda_{\rm dec}(\theta).
\end{equation}
It is decomposable if and only if
$\lambda\ge\lambda_{\rm dec}(\theta)$.
\end{theorem}

\begin{proof}
By Theorem~\ref{thm:DEC}, the map \(\Phi_{\lambda,\theta}\) is
decomposable precisely when \(F_{\lambda,\theta}(r)\ge0\) for all
\(r\ge0\), where
\begin{equation}
\label{eq:F}
F_{\lambda,\theta}(r)
=1+2r\sqrt{\lambda(1+\lambda)}
-2\min\{r,1\}-2\lambda h_r(\delta(\theta)).
\end{equation}
For \(q(\theta)<1\), we first seek the positive value of
\(\lambda\) for which \(F_{\lambda,\theta}(1)=0\), namely
\begin{equation}
 -1+2\sqrt{\lambda(1+\lambda)}-2\lambda q(\theta)=0.
\end{equation}
Its unique positive solution is the finite branch of
\eqref{eq:lambda}. In Appendix~\ref{app:threshold} we show that
\(r=1\) is a global minimiser at this value and establish the
monotonicity in \(\lambda\) needed to prove both sides of the threshold.
For \(q(\theta)=1\), the same appendix shows that
\(F_{\lambda,\theta}(1)<0\) for every finite \(\lambda\), which
justifies the second branch of \eqref{eq:lambda}. This proves the theorem.
\end{proof}

For example,
\begin{equation}
 q(0)=\frac12,\qquad
 \lambda_{\rm dec}(0)=\frac13.
\end{equation}
The case $q(\theta)=1$ occurs precisely for
\begin{equation}
 \theta\equiv\frac{\pi}{3},\ \pi,\ \frac{5\pi}{3}\pmod{2\pi}.
\end{equation}

Recall that a positive map is atomic if it cannot be written as a sum of
a $2$-positive and a $2$-copositive map.  Here $\Psi$ is $2$-copositive
if $\Psi\circ T$ is $2$-positive.  Write $\mathcal P_2$ for the cone of
$2$-positive maps and $\mathrm{co}\mathcal P_2$ for its copositive analogue.
On $\M_3\to\M_3$, the theorem of Yang--Leung--Tang
\cite{YangLeungTang2016} implies that every $2$-positive map is
decomposable; the same holds for $2$-copositive maps by composition with
$T$. Since $\CP\subseteq\mathcal P_2$ and
$\coCP\subseteq\mathrm{co}\mathcal P_2$, we obtain
\begin{equation}
 \DEC=\CP+\coCP=\mathcal P_2+\mathrm{co}\mathcal P_2.
\end{equation}
Consequently, indecomposability and atomicity are equivalent in this
dimension. We therefore obtain:

\begin{corollary}
\label{cor:atomic}
For every finite $\lambda\ge0$ and every phase $\theta$,
$\Phi_{\lambda,\theta}$ is atomic if and only if
$0\le\lambda<\lambda_{\rm dec}(\theta)$.
\end{corollary}

\begin{figure}[ht]
\centering
\includegraphics[width=0.62\textwidth]{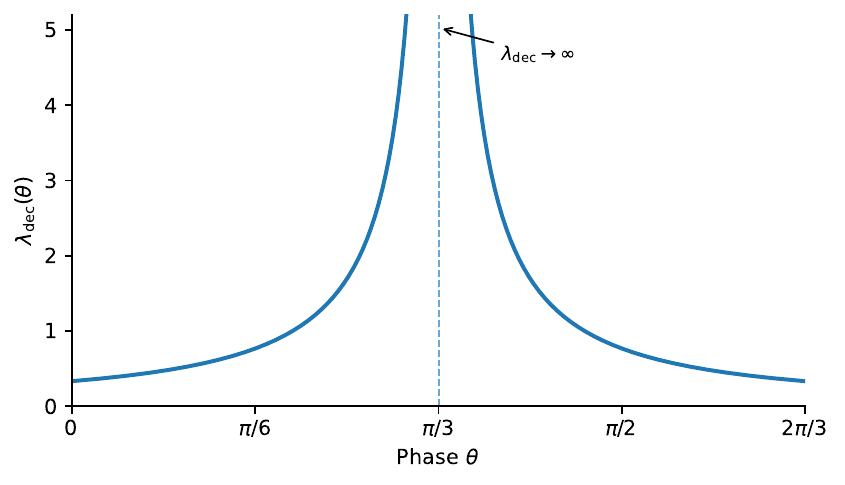}
\caption{Exact decomposability threshold $\lambda_{\rm dec}(\theta)$ over one
fundamental period. Maps below the curve are indecomposable (hence atomic),
while maps above it are decomposable. At $\theta=\pi/3$ the threshold
has the value $+\infty$, as in \eqref{eq:lambda}.}
\label{fig:threshold}
\end{figure}

Since $q(\theta)$ is $2\pi/3$-periodic, it is enough to consider one
fundamental interval.  Figure~\ref{fig:threshold} shows the corresponding
decomposability threshold.

\FloatBarrier

\section{PPT edge states and entanglement detection}
\label{sec:edge}

Since the maps \eqref{eq:phasefamily} are positive, their Choi
matrices are block-positive. In the indecomposable region of
Theorem~\ref{thm:threshold}, they are not positive semidefinite:
otherwise the maps would be completely positive. Thus they define
indecomposable entanglement witnesses. We now construct PPT edge
states of birank $(5,5)$ detected by these witnesses throughout
this region. The construction uses the invariant PPT operators
from Sec.~\ref{sec:dec}. Related low-rank PPT edge states have
been studied in
\cite{LewensteinKrausCiracHorodecki2000,HaKye2005,Clarisse2006,
Gulati2026RankBased}.

For $0<\alpha<\pi/3$, $t>0$ and $k\in\{0,1,2\}$, put
\[
 A_\alpha=\cos\alpha+\sqrt3\sin\alpha,\quad
 U_\alpha=e^{i\alpha},\quad V_{\alpha,k}=\omega^k e^{i\alpha}
\]
and define the two-qutrit state
\begin{equation}
\label{eq:edgestate}
 \sigma_{\alpha,t}^{(k)}
 :=\frac{\rho(A_\alpha,t,t^{-1};U_\alpha,V_{\alpha,k})}
 {3(A_\alpha+t+t^{-1})}.
\end{equation}

\Needspace{11\baselineskip}
In the standard product basis used in \eqref{eq:choi-matrix-full}, this state is
\begingroup
\setlength{\arraycolsep}{2.7pt}
\renewcommand{\arraystretch}{1.08}
\begin{equation}
\label{eq:edgestate-full}
\sigma_{\alpha,t}^{(k)}=\frac{1}{3(A_\alpha+t+t^{-1})}
\left(\begin{array}{ccc|ccc|ccc}
A_\alpha&0&0&0&U_\alpha&0&0&0&\bar U_\alpha\\
0&t^{-1}&0&V_{\alpha,k}&0&0&0&0&0\\
0&0&t&0&0&0&\bar V_{\alpha,k}&0&0\\ \hline
0&\bar V_{\alpha,k}&0&t&0&0&0&0&0\\
\bar U_\alpha&0&0&0&A_\alpha&0&0&0&U_\alpha\\
0&0&0&0&0&t^{-1}&0&V_{\alpha,k}&0\\ \hline
0&0&V_{\alpha,k}&0&0&0&t^{-1}&0&0\\
0&0&0&0&0&\bar V_{\alpha,k}&0&t&0\\
U_\alpha&0&0&0&\bar U_\alpha&0&0&0&A_\alpha
\end{array}\right).
\end{equation}
\endgroup
\begin{proposition}
\label{prop:edge}
For $0<\alpha<\pi/3$, $t>0$ and $k\in\{0,1,2\}$, the state
$\sigma_{\alpha,t}^{(k)}$ is a PPT entangled edge state, with
\begin{equation}
\label{eq:birank}
 \left(\operatorname{rank}\sigma_{\alpha,t}^{(k)},
 \operatorname{rank}\left[(\sigma_{\alpha,t}^{(k)})^\Gamma\right]\right)=(5,5).
\end{equation}
\end{proposition}

\begin{proof}
The trace of the numerator in \eqref{eq:edgestate} is
$3(A_\alpha+t+t^{-1})$, so the state is normalised.
The block eigenvalues give positivity and birank $(5,5)$.
In particular, each $2\times2$ block has rank one because
$t\,t^{-1}=|U_\alpha|^2=|V_{\alpha,k}|^2=1$.
The range-criterion proof of the edge property is given in
Appendix~\ref{app:edge}.
\end{proof}

\begin{proposition}
\label{prop:detect}
For every $\lambda\ge0$ and phase $\theta$, the map
$\Phi_{\lambda,\theta}$ is indecomposable if and only if at least one
state \eqref{eq:edgestate} satisfies
\begin{equation}
\label{eq:edge-detect}
 \operatorname{Tr}\!\left(C_{\Phi_{\lambda,\theta}}
 \sigma_{\alpha,t}^{(k)}\right)<0.
\end{equation}

Equivalently, this condition is

\begin{equation}
\label{eq:edge-ineq}
 \lambda t+\frac{1+\lambda}{t}-\cos\alpha
 +\sqrt3\sin\alpha
 +2\lambda\cos\!\left(\alpha+\frac{2\pi k}{3}-\theta\right)<0.
\end{equation}
\end{proposition}

\begin{proof}
By \eqref{eq:pairing}, the expectation in \eqref{eq:edge-detect}
has the sign of the left-hand side of \eqref{eq:edge-ineq}.
For $\lambda>0$, minimising in $t$ gives
$t=\sqrt{(1+\lambda)/\lambda}$.  Choose $k$ so that
$\cos(2\pi k/3-\theta)=-q(\theta)$.  In the limit
$\alpha\downarrow0$, the left-hand side of \eqref{eq:edge-ineq}
then tends to
\begin{equation}
 2\sqrt{\lambda(1+\lambda)}-1-2\lambda q(\theta)
 =F_{\lambda,\theta}(1).
\end{equation}
This is strictly negative throughout the indecomposable region by
Theorem~\ref{thm:threshold}, so it remains negative for sufficiently
small $\alpha>0$.  For $\lambda=0$, choose first $t>1$ and then
$\alpha>0$ sufficiently small.
Conversely, a decomposable witness is non-negative on every PPT state.
\end{proof}

The construction gives explicit PPT states for the exact threshold
\eqref{eq:lambda}. It does not require the state family to be
inequivalent to earlier families of low-rank PPT states.
Other cyclic two-qutrit states have been studied in
\cite{ZhangAliMa2026}, with a focus on their evolution under dephasing.

\section{Conclusions}
\label{sec:conclusions}

We obtained a necessary and sufficient decomposability criterion
for the entire seven-parameter class of Hermiticity-preserving
cyclic-sign covariant maps on $M_3$. It requires no positivity
assumption and gives an exact indecomposability test whenever the
map is positive. For a phase-dependent positive deformation of
the Choi map, we determined the indecomposable and atomic regions
exactly. The corresponding Choi matrices are indecomposable
entanglement witnesses, and we constructed PPT edge states
of birank $(5,5)$ detected by these witnesses throughout
the indecomposable region.

A possible direction for further work is to seek similar exact
decomposability criteria and PPT-state constructions in higher
dimensions. The $A_4$ symmetry and the block decomposition used here
are specific to $d=3$, so this would require a different symmetry
group and a new analysis of its invariant operators. Moreover, the
equivalence between indecomposability and atomicity used here does
not extend automatically to higher dimensions; these properties
would need to be studied separately.

\appendix

\section{Geometry of the function \texorpdfstring{$m_r$}{m\_r}}
\label{app:geometry}

\begin{proof}[Proof of Proposition~\ref{prop:m}]
For $z\neq0$ write
\begin{equation}
 w=-\frac{z}{|z|}.
\end{equation}
Then
\begin{equation}
 m_r(z)
 =
 -|z|\max_{\xi\in K_r}\operatorname{Re}(\bar w\xi),
 \qquad
 K_r=\Delta\cap r\D.
\end{equation}
The equilateral triangle $\Delta$ is invariant both under rotations by
$2\pi/3$ and under complex conjugation. Using these dihedral symmetries,
we may assume $w=e^{i\delta}$ with $0\le\delta\le\pi/3$; the nearest
vertex to $w$ is then $1$.

For $0\le r\le r_0(\delta)$, the point $rw$ lies in $\Delta$ and
maximises $\operatorname{Re}(\bar w\xi)$.  Thus

\begin{equation}
 h_r(\delta)=r.
\end{equation}

For $r_0(\delta)\le r\le1$, let $\xi$ be the intersection of
$|\xi|=r$ with the side of $\Delta$ joining $1$ and $\omega$,
chosen closer to $1$. The support direction $w=e^{i\delta}$ lies
between the outward normal of this side and the radial direction of
$\xi$. It is therefore a non-negative linear combination of these
two normals. Since both the side and the disk support their respective
convex sets at $\xi$, this point maximises the scalar product
over $\Delta\cap r\D$.
This side is at distance $1/2$ from
the origin.  Its outward normal
has angle $\pi/3$, and the tangent toward $1$ has angle $-\pi/6$.
The component along the normal is therefore $1/2$, while the
tangent component is $\sqrt{r^2-1/4}$.  Taking the scalar product
with $w=e^{i\delta}$ gives
\begin{equation}
 h_r(\delta)
 =
 \frac12\cos\left(\frac{\pi}{3}-\delta\right)
 +\sqrt{r^2-\frac14}\cos\left(\delta+\frac{\pi}{6}\right).
\end{equation}

For $r\ge1$, the disk contains all of $\Delta$.  The maximum is
attained at the nearest vertex, so

\begin{equation}
 h_r(\delta)=\cos\delta.
\end{equation}
\end{proof}

The following property is used in Appendix~\ref{app:threshold}.

\begin{lemma}
\label{lem:hconcave}
For each fixed $\delta\in[0,\pi/3]$, the function
$r\mapsto h_r(\delta)$ is non-decreasing and concave on $[0,\infty)$.
\end{lemma}
\begin{proof}
Let $w$ be the unit vector in the support direction, so that
\begin{equation}
 h_r(\delta)=\max_{\xi\in\Delta\cap r\mathbb D}
 \operatorname{Re}(\bar w\xi).
\end{equation}
Choose maximisers $\xi_i\in\Delta\cap r_i\mathbb D$,
$i=1,2$.  For $0\le t\le1$, convexity of $\Delta$ and the triangle
inequality imply
\begin{equation}
 t\xi_1+(1-t)\xi_2
 \in\Delta\cap(tr_1+(1-t)r_2)\mathbb D.
\end{equation}
Taking the real part of the scalar product with $w$ gives
\begin{equation}
 h_{tr_1+(1-t)r_2}(\delta)
 \ge t h_{r_1}(\delta)+(1-t)h_{r_2}(\delta).
\end{equation}
Monotonicity follows from the inclusion of the feasible sets when $r$
increases.
\end{proof}

\section{Proof of the threshold}
\label{app:threshold}

\begin{proof}[Details of Theorem~\ref{thm:threshold}]
Set $p(\lambda)=\sqrt{\lambda(1+\lambda)}$ and $q=\cos\delta$.
Assume first that $q<1$.  Let $\lambda_*>0$ be the unique positive
solution of $F_{\lambda,\theta}(1)=0$, namely
\begin{equation}
\label{eq:thresholdcandidate}
 p(\lambda_*)=\frac12+\lambda_*q.
\end{equation}
Indeed, squaring gives
\begin{equation}
 4(1-q^2)\lambda_*^2+4(1-q)\lambda_*-1=0,
\end{equation}
whose unique positive root is \eqref{eq:lambda}.  At this stage
$\lambda_*$ is only a candidate for the decomposability threshold.

We first show that $r\mapsto F_{\lambda_*,\theta}(r)$ is convex on
$[0,\infty)$.  By Lemma~\ref{lem:hconcave}, $h_r(\delta)$ is concave in $r$.
Moreover, $r\mapsto\min\{r,1\}$ is concave.
Thus \eqref{eq:F} is a sum of an affine function and two convex functions.

It remains to verify that $r=1$ is a minimiser.
The one-sided derivatives at $r=1$ are
\begin{align*}
 F'_{\lambda_*,\theta}(1^-)
 &=2p(\lambda_*)-2-2\lambda_*
 \left(q-\frac{\sin\delta}{\sqrt3}\right)
 =-1+\frac{2\lambda_*\sin\delta}{\sqrt3},\\
 F'_{\lambda_*,\theta}(1^+)&=2p(\lambda_*)>0.
\end{align*}
Using \eqref{eq:thresholdcandidate} one finds
\begin{equation}
 4\lambda_*^2\sin^2\delta
 =1-4(1-q)\lambda_*<1.
\end{equation}
Hence $2\lambda_*\sin\delta<1$, and in particular
$F'_{\lambda_*,\theta}(1^-)<0$.
Convexity now implies that $r=1$ is a global minimiser.  Since
$F_{\lambda_*,\theta}(1)=0$, we conclude that
\begin{equation}
 F_{\lambda_*,\theta}(r)\ge0\qquad (r\ge0).
\end{equation}

For every fixed $r>0$ and $\lambda>0$,
\begin{equation}
 \frac{\partial F_{\lambda,\theta}(r)}{\partial\lambda}
 =2r\,\frac{1+2\lambda}{2\sqrt{\lambda(1+\lambda)}}
 -2h_r(\delta)>0,
\end{equation}
since $(1+2\lambda)/(2\sqrt{\lambda(1+\lambda)})>1$ and
$h_r(\delta)\le r$.  At $r=0$, $F_{\lambda,\theta}(0)=1$.
By continuity, the same monotonicity comparison applies down to
$\lambda=0$.  Therefore $F_{\lambda,\theta}(r)\ge0$ for all $r$
when $\lambda\ge\lambda_*$; for $\lambda<\lambda_*$ we have
$F_{\lambda,\theta}(1)<0$.  This proves the threshold for $q<1$.

Finally, if $q=1$, then for every finite $\lambda\ge0$,
\begin{equation}
 F_{\lambda,\theta}(1)
 =-1+2\sqrt{\lambda(1+\lambda)}-2\lambda<0,
\end{equation}
because $\sqrt{\lambda(1+\lambda)}<\lambda+\tfrac12$.
Hence the map remains indecomposable for every finite $\lambda$,
justifying the value $+\infty$ in \eqref{eq:lambda}.
\end{proof}

\section{Range criterion for the phase-adapted states}
\label{app:edge}

\begin{proof}[Details of Proposition~\ref{prop:edge}]
The full matrix of the state is given in \eqref{eq:edgestate-full}.
In the basis $|11\rangle,|22\rangle,|33\rangle$, the $3\times3$
block of $\sigma_{\alpha,t}^{(k)}$ (before normalisation) has eigenvalues
\begin{equation}
 0,\quad 2\sqrt3\sin\alpha,\quad
 3\cos\alpha+\sqrt3\sin\alpha.
\end{equation}
The corresponding block after partial transposition has the same
eigenvalues.
Moreover $t\cdot t^{-1}=|U_\alpha|^2=|V_{\alpha,k}|^2=1$, so the
three $2\times2$ blocks of both $\sigma$ and $\sigma^\Gamma$ have
rank one.  Thus both operators are positive with birank $(5,5)$.

We use the range characterization of PPT edge states
\cite{LewensteinKrausCiracHorodecki2000}: there is no nonzero
product vector $x\otimes y\in\operatorname{Ran}\sigma$ such that
$\bar x\otimes y\in\operatorname{Ran}\sigma^\Gamma$.
We now check this condition.
  Write $\omega=e^{2\pi i/3}$, label indices
$i=1,2,3$ cyclically, and set $V=\omega^k e^{i\alpha}$ and
$U=e^{i\alpha}$.  The diagonal-block kernels are spanned by the vectors
\begin{equation}
 (1,\omega,\omega^2),\qquad
 (1,\omega^{k-1},\omega^{2(k-1)})
\end{equation}
for $\sigma$ and $\sigma^\Gamma$, respectively.
Orthogonality of the product vectors to these kernels gives
\begin{equation}
\label{eq:edge-diag}
 \sum_{i=1}^3\omega^{-(i-1)}x_i y_i=0,
 \qquad
 \sum_{i=1}^3\omega^{-(k-1)(i-1)}\bar x_i y_i=0.
\end{equation}
The $2\times2$ kernels give, for the three cyclic pairs
$(i,j)=(1,2),(2,3),(3,1)$,
\begin{equation}
\label{eq:edge-offdiag}
 x_jy_i=t\bar V x_i y_j,
 \qquad
 \bar x_jy_i=tU\bar x_i y_j.
\end{equation}

Let $S_x=\{i:x_i\ne0\}$ and $S_y=\{i:y_i\ne0\}$.
The first equation in \eqref{eq:edge-offdiag} shows that
$x_i y_j$ vanishes if and only if $x_j y_i$ vanishes for $i\ne j$.
Indeed, if $i\in S_x\setminus S_y$, choose $j\in S_y$; then
$x_i y_j\ne0$ but $x_j y_i=0$, a contradiction. By symmetry
$S_x=S_y=:S$.
If $|S|=1$, the first equation in \eqref{eq:edge-diag} is impossible.

If $|S|=3$, division of the two equations in
\eqref{eq:edge-offdiag} shows that, for each cyclic pair,
\begin{equation}
 \frac{x_j/\bar x_j}{x_i/\bar x_i}
 =\frac{\bar V}{U}=\omega^{-k}e^{-2i\alpha}.
\end{equation}
Multiplication over all three pairs yields
$1=e^{-6i\alpha}$, impossible for $0<\alpha<\pi/3$.

Finally suppose $|S|=2$.  Write its indices as $(i,j)$, with $j$
cyclically following $i$ (thus $j-i\equiv1\pmod3$).
The two conditions in \eqref{eq:edge-diag} then imply, respectively,
\begin{equation}
 \frac{x_jy_j}{x_iy_i}=-\omega^{j-i},
 \qquad
 \frac{\bar x_jy_j}{\bar x_iy_i}=-\omega^{(k-1)(j-i)}.
\end{equation}
Dividing these identities gives
\begin{equation}
 \frac{x_j/\bar x_j}{x_i/\bar x_i}=\omega^{2-k}.
\end{equation}
On the other hand, \eqref{eq:edge-offdiag} yields
$\omega^{-k}e^{-2i\alpha}$.  Comparison forces
$e^{-2i\alpha}=\omega^2$, impossible for $0<\alpha<\pi/3$.
Thus no such product vector exists.
The state is therefore PPT edge, and in particular PPT entangled.
\end{proof}

\bibliographystyle{apsrev4-2}
\bibliography{A4_ar}

\end{document}